\documentclass{ws-mpla}
\usepackage[english]{babel}
\usepackage[tracking,spacing,kerning,letterspace=0,babel]{microtype}
\begin{document}

\markboth{Serge Duarte Pinto}
{GEM applications outside high energy physics}

\catchline{}{}{}{}{}

\title{\uppercase{GEM applications outside high energy physics}}

\author{\footnotesize \uppercase{Serge Duarte Pinto} }

\address{CERN, 1211 Geneva 23, Switzerland\\
Serge.Duarte.Pinto@cern.ch}

\maketitle

\pub{Received (Day Month Year)}{Revised (Day Month Year)}

\begin{abstract}
From its invention in 1997, the Gas Electron Multiplier has been applied in nuclear and high energy physics experiments.
Over time however, other applications have also exploited the favorable properties of \textsc{gem}s.
The use of \textsc{gem}s in these applications will be explained in principle and practice.

This paper reviews applications in research, beam instrumentation and homeland security. The detectors described measure neutral radiations such as photons, x-rays, gamma rays and neutrons, as well as all kinds of charged radiation.
This paper provides an overview of the still expanding range of possibilities of this versatile detector concept.
\keywords{Gas Electron Multipliers; GEMs; X-rays; Neutrons; Beam Instrumentation; Spherical Detectors.}
\end{abstract}

\ccode{PACS Nos.: include PACS Nos.}

\section{Introduction}	
The Gas Electron Multiplier (\textsc{gem}) has been applied in many high-energy physics experiments, particularly in the zones with the highest particle rate.
But \textsc{gem}s have many interesting properties besides their superb rate capability.
These features have been exploited in applications in other areas of research, and also outside the scientific domain.

The natural suppression of ion backflow and photon feedback makes \textsc{gem}s suitable for photodetectors.
A \textsc{gem} foil itself can support a solid photoconverter, while also acting as a gas amplification stage.
Efficient and cost-effective photodetectors can thus be made with virtually unlimited size and without dead zones.
The readout pattern can be tailored to the requirements of the application in terms of spatial resolution and geometry.
The principle of using a \textsc{gem} foil as a substrate for a converter layer has been applied to other types of neutral radiation as well; examples will be shown of \textsc{gem}-based detectors for neutrons, hard x-rays and gamma rays.

For detection of x-rays or neutrons, the gas can be used as a converter as well.
The recent development of a spherically shaped \textsc{gem} eliminates parallax errors associated with this method when applied to diffraction measurements or pinhole imaging.
The principle will be explained.

\section{Neutral Radiation}
\begin{figure}
\includegraphics[width=\columnwidth]{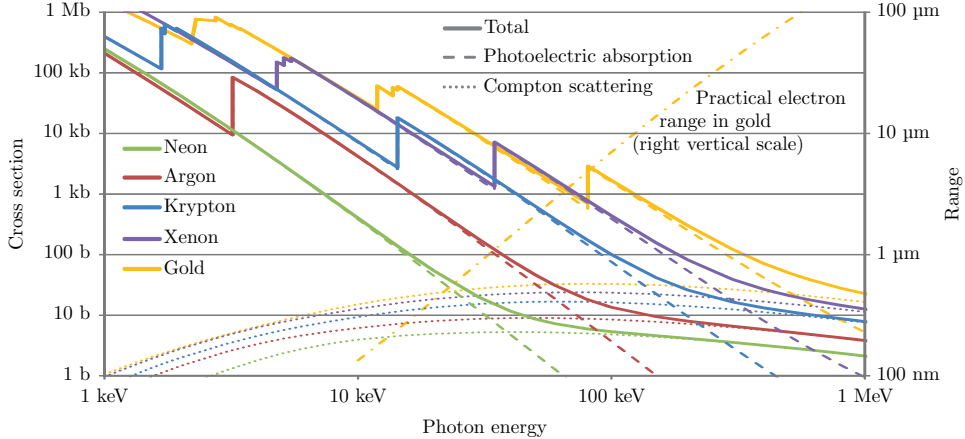}
\caption{Cross sections of selected elements used as x-ray converters in gaseous detectors. Dashed and dotted curves decompose the total interaction cross sections into photoelectric absorption (dominant for soft x-rays) and Compton scattering (significant for harder x-rays).
Other interactions have negligible cross sections in this energy range.
For gold converters the \emph{practical range} (as defined in~[\protect\refcite{PracticalRange}]) of the photoelectrons is also plotted.}
\label{CrossSections_x-rays}
\end{figure}

\begin{figure}
\includegraphics[width=\columnwidth]{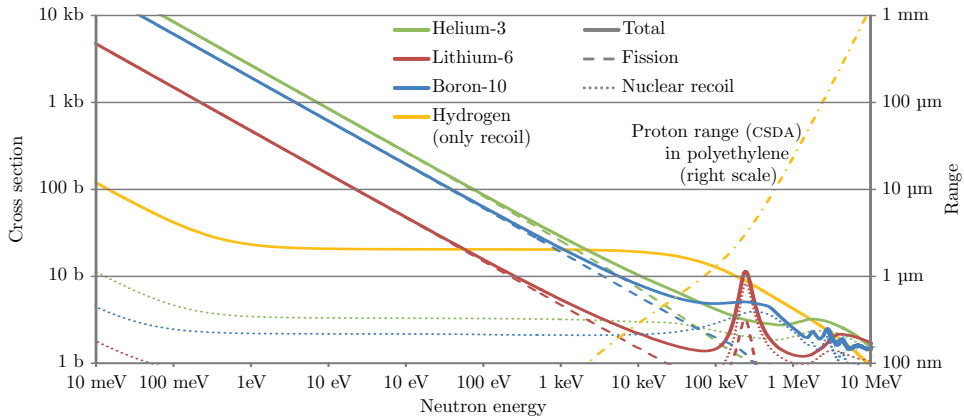}
\caption{Cross sections of selected isotopes used as neutron converters in gaseous detectors. Dashed and dotted curves decompose the total interaction cross sections into fission (dominant for thermal neutrons) and nuclear recoil reactions (significant for fast neutrons).
Other interactions have negligible cross sections in this energy range.
Hydrogen does not undergo fission reactions and the displayed cross section is only for nuclear recoils.
As hydrogen converters are often applied in the form of hydrogen-rich plastic foils (polyethylene or polypropylene), the proton range (calculated with the continuous slowing down approximation (\textsc{csda})) in polyethylene is also shown as a function of energy.}
\label{CrossSections_neutrons}
\end{figure}

Detection of neutral radiation such as photons and neutrons relies on a conversion mechanism that yields charges in the drift volume of the detector.
In the case of photons the conversion mechanism is normally the photoelectric effect, although for x-rays of sufficiently high energy ($\ge100$ keV) Compton scattering could also play a role.
For thermal neutrons, conversion takes place via a nuclear fission reaction with an appropriate isotope, while higher energy neutrons may cause nuclear recoils.
Figure~\ref{CrossSections_x-rays} shows interaction cross sections of a selection of converter materials for x-rays, and figure~\ref{CrossSections_neutrons} for neutrons, both as a function of energy of the incoming particle.
Apart from structures that are specific for each converter, the cross sections generally decrease with increasing energy $E$.
For x-rays the decrease is particularly steep: $\sigma\propto\frac{Z^4}{AE^3}$ (with $Z$ and $A$ atomic number and mass number of the absorber), while for thermal neutrons (meV range) it is inversely proportional to their velocity, or $\sigma\propto\frac{1}{\sqrt{E}}$.

How a converter of neutral radiation is implemented in a gaseous detector depends on the requirements in terms of efficiency and space, time and energy resolution.
Very broadly, three approaches have emerged with the converter in solid, liquid or gaseous phase.
Their respective merits and limitations, along with some examples, will be discussed below.

\subsection{Gaseous Converters}
A gas-phase converter is often an attractive choice if the gas is also suitable as a gas amplification medium, optimally a noble gas.
The main technical challenge of using a gaseous converter is the low density compared to solid or liquid phases, and thus the low stopping power.
Nevertheless, gaseous converters can potentially provide the highest detection efficiencies since no primary charge is lost in the transfer mechanism.
Figures~\ref{CrossSections_x-rays} and \ref{CrossSections_neutrons} show that noble gases exist with cross sections of tens or even hundreds of kilobarns, be it for the lowest energies of x-rays and neutrons.
Gases like Xe and $^3$He are also very costly, mandating either a recirculation system or an ultra-high vacuum compatible sealed detector.

For reasons of efficiency, gaseous converters are often used at high pressures or with a wide drift gap.
Operation of \textsc{gem}s at a high gas pressure can be challenging; the maximum attainable gain tends to decrease with increasing pressure (though less so for He), while the the working voltages roughly scale with the pressure.\cite{XrayGas1,XrayGas2}
Another method to reach full efficiency is shown in Ref.~\refcite{WAXS}; a triple \textsc{gem} detector with a 1D (strip) readout, where the \textsc{gem}s are ``lied down'' as in a time projection chamber.
This way the x-rays traverse the drift region sideways, and the detector is designed to be several attenuation lengths wide to reach the desired efficiency.

\begin{figure}
\includegraphics[width=\columnwidth]{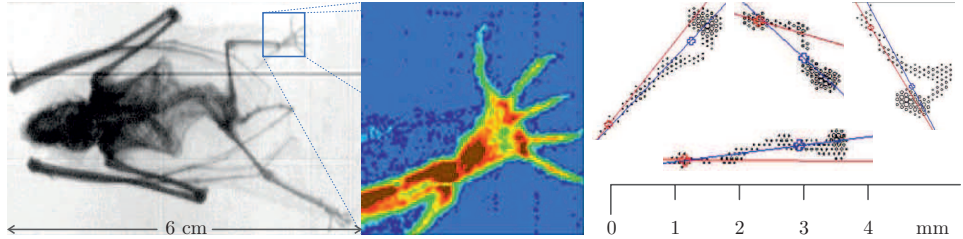}
\caption{Left: absorption radiography of a small bat, made with a double \textsc{gem} detector and an $^{55}$Fe source.\protect\cite{Bat} The color-scaled frame in the center reveals the attainable level of detail. Right: space-resolved 5.4 keV x-ray conversions in a Ne/\textsc{dme} (dimethyl ether) 80/20\% gas mixture.
The reconstruction is shown, marking the conversion point with a red cross and the direction of emission of the photoelectron with a red line.\protect\cite{Polarimeter}}
\label{x-rayGEM}
\end{figure}

Efficiency is not always a concern, and \textsc{gem}-based x-ray and neutron detectors offer high count rate capability over large areas with a good spatial resolution.
Figure~\ref{x-rayGEM} (left) shows an absorption radiography image made with a small double \textsc{gem} prototype and a simple $^{55}$Fe x-ray source (5.9 keV).
On the right side of the figure are projections of x-ray conversions in a Ne/\textsc{dme} 80/20\% gas mixture.
The images on the right are made with a fine-pitch single \textsc{gem} (90 $\mu$m between holes) read out by a pixel chip, each \textsc{gem} hole aligned with a readout pixel.\protect\cite{Polarimeter}
By resolving the conversion events spatially one can reconstruct the direction of emission of the primary photoelectron (indicated in the figure), which is related to the polarization axis of the incoming photon.

The best gaseous neutron converter is $^3$He, with a cross section that is very large for neutron interactions and negligible for gamma interactions.
It had been used with many kinds of gaseous detectors, including \textsc{gem}s.\cite{He-3}
In contrast to the heavier noble gases used for x-ray detection, $^3$He can be used at very high pressures with a triple \textsc{gem}.
Another converter is $^{10}$BF$_3$, an effective quencher and therefore not suitable for high pressures.

\begin{figure}[b]
\includegraphics[width=\columnwidth]{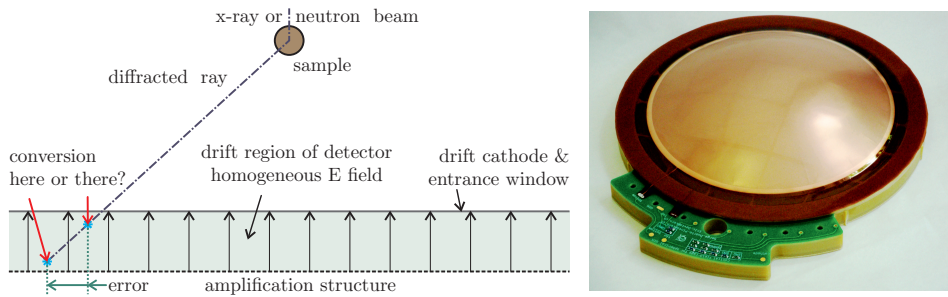}
\caption{Left: the cause of parallax error in an x-ray or neutron diffraction detector with a gaseous converter.
Right: the first spherical \textsc{gem} assembly designed to eliminate this error.}
\label{Spherical}
\end{figure}

In neutron or x-ray diffraction studies, where the radiation to be detected comes from a point-like source (the sample), an issue arises with planar gas detectors: the parallax error.
This is the error in the position reconstruction caused by the spread in depth of conversion of x-rays or neutrons in the drift region, see Figure~\ref{Spherical}.
This problem can be somewhat alleviated by using high cross section and high pressure converters, or a specially shaped entrance window.
A spherical \textsc{gem} detector eliminates it entirely, since radiation always enters the detector in the same direction as the drift field.
Spherical \textsc{gem}s are difficult to fabricate,\cite{Spherical1} but promise the ultimate diffraction imaging performance.\cite{Spherical2,Spherical3}

\subsection{Solid Converters}
The most familiar solid converters are the photocathodes used since long in vacuum photomultipliers.
The photocathode used most often with gas detectors is CsI,\cite{CsI} a monoalkali which is only sensitive to ultraviolet light ($\ge 6.2$ eV).
It is vacuum evaporated on a clean and non-reactive substrate, in practice often gold-plated copper.
CsI is much less vulnerable than bialkali photocathodes to impurities in the gas (notably oxygen and water) and ion feedback.
CsI converters can be applied to the drift electrode (semitransparent photocathode),\cite{CsICathode} or directly to the top electrode of the first \textsc{gem} (reflective photocathode).\cite{CsIGEM1,CsIGEM2}
The quantum efficiency of a CsI photocathode in a gaseous detector is limited by backscattering of the photoelectron by gas molecules.
To suppress backscattering most CsI-based detectors are operated in gas mixtures rich in CH$_4$ (methane)\cite{Backscattering1} or CF$_4$,\cite{Backscattering2} both known to absorb kinetic energy from electrons very effectively.
CsI-based gaseous photodetectors are attractive for applications that demand large area coverage with minimal dead space, such as Cherenkov detectors.
A notable example of this is a triple \textsc{gem} photodetector developed for the \textsc{phenix} experiment at \textsc{rhic}.\cite{HBD}
It is operated with CF$_4$ , which also serves as the Cherenkov radiator, thus obviating the need for a \textsc{vuv}-transparent window and maximizing photon yield.
It also runs on a slightly reversed drift field, to suppress the otherwise severe background of charged particles traversing the detector as well.

Visual range bialkali photocathodes have also been pursued with \textsc{gem}s.
These photocathodes are chemically reactive, and cannot be exposed even briefly to air; sub-ppm level impurities of oxygen or water in the gas dramatically reduce the lifetime of the photocathode.
In addition, they are very sensitive to photon- and ion-induced secondary effects.\cite{GPMs}
The former issue can be alleviated by coating the photocathode with thin films (few hundred \AA) of NaI or CsI, but such protected photocathodes have quantum efficiencies of a few percent at most.\cite{ProtectiveFilm}
Alternatively, the detector must be made ultra-high vacuum compatible by sealing and choice of materials.
Regarding the ion and photon feedback issues, \textsc{gem}s have rather benign properties.
Photon feedback is suppressed because the gas avalanches take place inside the narrow \textsc{gem} holes, and most avalanche photons are produced in the last \textsc{gem} of a cascade, these photons are absorbed by the \textsc{gem}s or the readout board.
\textsc{Gem}s also suppress ion feedback by attracting ions to the various electrodes before they can flow all the way back to the photocathode; depending on operational parameters, roughly 99\% of ions are intercepted in this way.

Solid converters exist also for neutrons and even for x-rays.
For thermal neutrons $^6$Li and especially $^{10}$B can be used, applied to the cathode and the \textsc{gem} electrodes.
The fission products are an $\alpha$ particle of $\sim2$ MeV and either a triton (for $^6$Li) or a lithium ion (for $^{10}$B), isotropically emitted in opposite directions.
The Li ion, due to its low energy and high charge state, will rarely escape the converter; therefore many conversions do not lead to detection.
To have at least the $\alpha$ escape the solid, the layer thickness is optimized to $\sim2.4 \mu$m.\cite{B-10}
One such layer only has an efficiency of a few percent, so many detectors use multiple boron-coated \textsc{gem}s operating with a gain close to unity, and only the last \textsc{gem} amplifying.\cite{B-10_2,CASCADE}

A similar tactic of having cascaded unity gain stages of converter \textsc{gem} foils is used in Ref.~\refcite{Gold} with a 3 $\mu$m gold layer as a converter for high energy x-rays.
Figure~\ref{CrossSections_x-rays} shows that gold has a high cross section, and that from about 100 keV photon energy the photoelectron has a good chance of escaping the gold layer (right scale).
This detector has an efficiency of around a percent at 141 keV.

Higher energy neutrons are also difficult to detect efficiently; beyond 10 keV all fission cross sections are below 10 barn.
At such energies normally hydrogen-rich materials (paraffine plates, polyethylene foils) are used to provoke nuclear recoil.
The recoiling proton has an energy $E_p=E_ncos^2\theta$, with $E_n$ the neutron energy and $\theta$ the recoil angle.
Figure~\ref{CrossSections_neutrons} (to the right scale) shows the \textsc{csda} range of recoil protons from head-on neutron collisions ($\theta=0$).
For high-flux neutron sources like reactors, detectors do not need to be efficient as long as they are insensitive to gamma rays.\cite{nGEM,polyethylene}

\subsection{Liquid Converters}
Liquid converters are used in cryogenic two-phase detectors for radiation that interacts only by the weak force (neutrinos, \textsc{wimp}s), and therefore needs a lot of mass to be detected at all.
They are also proposed for hard x-ray applications where detection efficiency is paramount, such as positron emission tomography.
To maximize the mass, and minimize the probability of primary electron capture, high purity, heavy noble liquids are used: Ar, Kr or Xe.
Ionization charge in the liquid can be extracted to the gas by an electric field, and then amplified by \textsc{gem}s.\cite{2Phase}
The operation of multiple \textsc{gem} cascades in the high gas densities in a cryostat can be so difficult that at a certain density the maximum gain drops below that of a single \textsc{gem}.\cite{HighPressure}
Solutions to this are sought in adding a quencher (e.g. CH$_4$) or using noble mixtures.
The necessity to work in a cryostat and the difficult operational conditions for a gas detector will probably prevent liquid converters from becoming a mainstream solution for many applications.
But the combination of a large mass and a high sensitivity to weakly ionizing conversions makes this concept uniquely suited to rare event applications such as dark matter searches.

\section{Beam Instrumentation}
\begin{figure}
\includegraphics[width=\columnwidth]{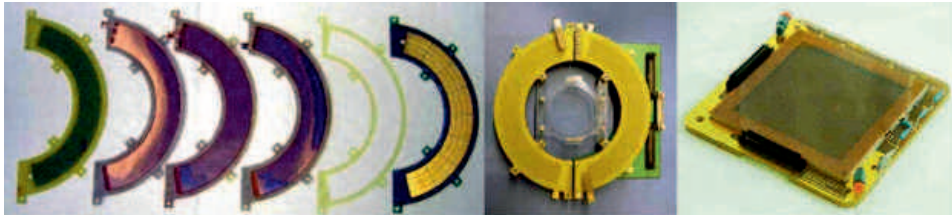}
\caption{Left: exploded and assembled view of a triple \textsc{gem} Bhabha luminometer, shaped to embrace the beam pipe.\protect\cite{BIFrascati} Right: ultra-light single \textsc{gem} detector developed for the antiproton decelerator at \textsc{cern}.\protect\cite{AD}}
\label{BI}
\end{figure}
\textsc{Gem}s have found various applications in beam instrumentation.
The \textsc{dafne} collider in Frascati used triple \textsc{gem} detectors for luminosity measurements.\cite{BIFrascati}
This is done either by counting photons emitted in forward direction from single bremsstrahlung events in the interaction point, or by tracking Bhabha-scattered electrons and positrons (see Figure~\ref{BI}, left); the rates of both these processes are proportional to the luminosity.
Some beam facilities use \textsc{gem} detectors as time projection chambers (\textsc{tpc}),\cite{BIFrascati,IonTPC} i.e. with the \textsc{gem} foils parallel to the beam axis and measuring one transverse profile directly while accessing the other projection by drift time measurement of the ionization.
Such \textsc{tpc} mode profilers are attractive as the beam crosses only thin windows and gas, but not \textsc{gem} foils and a readout structure; moreover one detector measures location and direction of each particle (imperative in the \textsc{fair} fragment separator, Ref.~\refcite{IonTPC}).
Single \textsc{gem} detectors are used as transverse profile monitors in the antiproton decelerator at \textsc{cern} (see Figure~\ref{BI}, right),\cite{AD} and triple \textsc{gem} detectors will be installed for the same purpose in all higher energy beam lines in the other experimental areas at CERN.

\section{Homeland Security}
In a development funded by the US Department of Homeland Security, large \textsc{gem}s are used to scan cargo for nuclear contraband.
The method of muon tomography uses cosmic ray muons as a probe to locate high-Z materials such as fissionable nuclei, by reconstructing scattering centers.\cite{MT}
Two stations of large area \textsc{gem} trackers are placed on all sides around a container, and incoming and outgoing muons are tracked; the point of closest approach is considered to be the scattering center.
A close concentration of scattering centers with wide scattering angles could indicate the presence of uranium or plutonium.
If the sample is exposed for long enough, also materials with lower Z such as lead and even iron blocks can be identified and their shapes reconstructed.

\section{Conclusion}
The detector concepts and applications described and cited above are only samples of the enormous amount of work done, and still being done, on development of gaseous detectors based on the gas electron multiplier.
In trying to give a most comprehensive overview of non-\textsc{hep gem} development I have only scratched the surface of this exciting field.
The reader is encouraged to look up the references and the references in those papers, for a more in-depth look at any specific application.

\end{document}